\documentclass[a4paper,10pt]{article}
\pdfoutput=1

\usepackage{jcappub}
\usepackage[utf8]{inputenc}
\usepackage{amsmath}
\usepackage{amsfonts}
\usepackage{amssymb}
\usepackage{graphicx}
\usepackage{url}
\usepackage{dcolumn}
\usepackage{subcaption}
\usepackage{caption}
\usepackage{float}
\usepackage{bm}
\usepackage{latexsym}
\usepackage{epsfig}
\usepackage[normalem]{ulem}
\usepackage{graphicx}
\usepackage[colorlinks=true]{hyperref}

\newcommand{\Ne}{n_{\rm e}}
\newcommand{\NH}{n_{\rm H}}

\newcommand{\op}{\omega_{\rm p}}

\newcommand{\me}{m_{\rm e}}

\newcommand{\ma}{m_{\rm a}}
\newcommand{\Dosc}{\Delta_{\rm osc}}

\newcommand{\De}{\Delta_{\rm e}}

\newcommand{\Da}{\Delta_{\rm a}}
\newcommand{\Dagax}{\Delta^x_{\rm \gamma a}}
\newcommand{\Dagay}{\Delta^y_{\rm \gamma a}}
\newcommand{\Daga}{\Delta_{\rm \gamma a }}

\newcommand{\gammaa}{{\rm \gamma a}}

\newcommand{\Rc}{R_{\rm C}}
\newcommand{\Rv}{R_{\rm V}}

\begin{document}
\title{Constraints on non-resonant photon-axion conversion from the Planck satellite data}
\author[a,b,c] {Suvodip Mukherjee,}\emailAdd{mukherje@iap.fr}
\author[d] {Rishi Khatri}\emailAdd{khatri@theory.tifr.res.in}
\author[a,b,c,e]{and Benjamin D. Wandelt}\emailAdd{bwandelt@iap.fr}
\affiliation[a]{Center for Computational Astrophysics, Flatiron Institute, 162 5th Avenue, 10010, New York, NY, USA}
\affiliation[b]{Institut d'Astrophysique de Paris\\ 98bis Boulevard Arago, 75014 Paris, France}
\affiliation[c]{Sorbonne Universites, Institut Lagrange de Paris \\ 98 bis Boulevard Arago, 75014 Paris, France}
\affiliation[d]{Department of Theoretical Physics, Tata Institute of Fundamental Research\\ Homi Bhabha Road, Mumbai, 400005, India}
\affiliation[e]{Dept. of Astrophysical Sciences, Princeton University, Princeton, NJ 08544, USA}
\date{\today}
\abstract{The non-resonant conversion of Cosmic Microwave Background (CMB) photons into scalar as well as light pseudoscalar particles such as axion-like particles (ALPs) in the presence of turbulent magnetic fields can cause a unique, spatially fluctuating  spectral distortion in the CMB. We use the publicly available Planck temperature maps for the frequency channels (70-545 GHz) to obtain the first ALP distortion map using $45\%$ clean part of the sky. The $95^{th}$ percentile upper limit on the RMS fluctuation of ALP distortions from the cleanest part of the CMB sky at $15$ arcmin angular resolution is $18.5 \times 10^{-6}$. The RMS fluctuation in the distortion map is also consistent with different combinations of frequency channels and sky-fractions.}
\maketitle
\section{{Introduction}}
Distortions in the blackbody spectrum of Cosmic Microwave Background (CMB) are expected from several physical effects like thermal Sunyaev-Zeldovich ($y$-type distortion) in clusters of galaxies, Silk damping, axions, recombination lines, dark-matter annihilations, dark matter decay etc. \cite{zeldovich,sz1970, Chluba:2011hw, Chluba:2012gq,
  Khatri:2012tv, Khatri:2012rt, Khatri:2012tw, sk2013,Chluba:2016bvg,
  Hill:2015tqa, Emami:2015xqa, Hasselfield:2013wf,Bleem:2014iim, Ade:2015gva, Staniszewski:2008ma, 2010A&A...518L..16Z, 2013A&A...550A.134P,Mukherjee:2018oeb}. These effects span a wide range of redshifts from $z=2\times 10^6$ to $z=0$ and are an excellent probe of both the early and the late time cosmic evolution. The spectral distortions can be spatially isotropic, affecting only the CMB monopole intensity, or anisotropic with spatial fluctuations in the sky. The CMB spectral distortions arising from photon-ALPs conversion can also be polarized due to the resonant conversion in the large scale coherent magnetic field such as that of Milky Way \cite{Mukherjee:2018oeb} and galaxy clusters.  {The non-resonant photon-ALPs conversion due to the small scale magnetic field of Milky Way, galaxy clusters and voids can produce unpolarized spectral distortion in CMB \cite{Mukherjee:2018oeb}.}
      \begin{figure}[H]
\centering
     \includegraphics[trim={0.5cm 0.cm 0.5cm 0.7cm}, clip, width=1.0\linewidth]{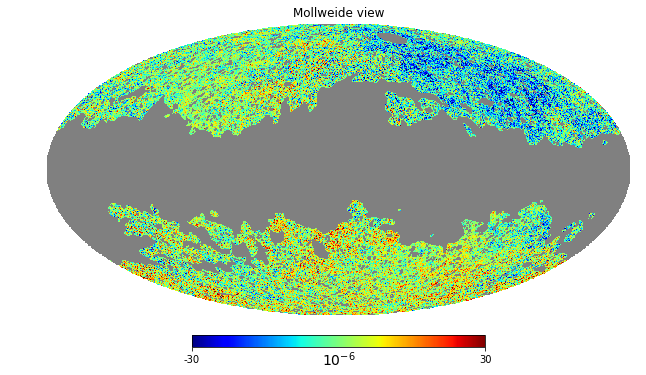}
        \caption{Internal Linear Combination Map of Axion Spectral Distortion ({ASD}) amplitude $P(\gamma\rightarrow  a)$ (See Eq. \eqref{spectrum}) for $f_{sky}= 0.45$ obtained from the six Planck-2015 temperature maps from 70 to 545 GHz. The details are provided in the results section.}\label{Fig:maphs}
     \end{figure}
  {The most stringent full sky observational constraints on the monopole part of the CMB spectral distortions come from the Cosmic Background Explorer-Far Infrared Absolute Spectrophotometer (COBE-FIRAS) \cite{firas,firasan,fm2002,2009ApJ...707..916F} with a $2\sigma$ upper limit on the $y$-type distortion of $y\le 15 \times 10^{-6}$ and on the $\mu$-type distortion of $\mu \le 9 \times 10^{-5}$. The Planck experiment with its multiple channels and wide frequency coverage, has allowed us for the first time after COBE to study and constrain other types of spectral distortions near the peak of the CMB spectrum. Since Planck, unlike COBE-FIRAS, does not have an absolute calibrator, we can only study the spectral distortions which are anisotropic. The spatially fluctuating $y$-type spectral distortions have been measured in  clusters \cite{Hill:2015tqa, Emami:2015xqa, Hasselfield:2013wf,Bleem:2014iim, Ade:2015gva, Staniszewski:2008ma, 2010A&A...518L..16Z, 2013A&A...550A.134P}, an upper and lower bound on the average distortions was obtained from the Planck data in \cite{Khatri:2015jxa} and an upper bound on the anisotropic $\mu$-type distortions from non-Gaussianity \cite{Pajer:2013oca} was obtained from the Planck data in \cite{Khatri:2015tla}.} 
 
 {In this paper, we study the spectral distortions that can originate from the non-resonant photon ALP conversion \cite{Mukherjee:2018oeb} (Axion Spectral Distortion (ASD)) in the presence of stochastic magnetic fields. This process can induce an unpolarized CMB spectral distortion signal that fluctuates  spatially and has a spectral shape that is different from the other known spectral distortions.   {This new scheme enables to probe the existence of ALPs in nature over a vast range of masses ($m_a < 10^{-11}$ eV). ALPs are potential candidates for cold dark matter and are predicted by string theory scenarios \cite{Arvanitaki:2009fg}. Several ground based experiments such CAST \cite{Anastassopoulos:2017ftl}, ALPS-II \cite{Bastidon:2015efa}, MADMAX \cite{Majorovits:2016yvk}, ADMX \cite{2010PhRvL.104d1301A}, CASPER \cite{2014PhRvX...4b1030B} are searching for ALPs. The method explored in this work is an independent method to detect ALPs using spatially varying spectral distortion signal in the blackbody spectrum of CMB.} Using the frequency spectrum of the  {ASD} signal, we obtain the  {ASD} sky map (shown in Fig \ref{Fig:maphs}) using the Internal Linear Combination (ILC) component separation method \cite{Tegmark:1995pn} on multi-frequency Planck sky maps (70- 545 GHz) smoothed to a common resolution of $15$ arcmin.}

 The  {ASD} signal depends strongly on the structure of the magnetic field and inhomogeneities in the electron density. We therefore need independent constraints on the $3-$D structure of electron density and magnetic fields to translate the bounds on the  {ASD} into constraints on Photon-ALPs coupling. We will use simple idealized models of magnetic field and electron density to derive joint bounds on the Photon-ALPs coupling, magnetic field and electron density.  The polarized spectral distortions in the polarized sky maps from Planck can be used to provide a bound on the photon-axion coupling strength in a narrow mass range of ALPs, $10^{-13}$ eV $\leq m_a \leq 10^{-11}$ eV. We leave the analysis of polarized  {ASD} for future work. 
\section{{Mechanism}}
The conversion of the photons ($A_x, A_y$) into axions ($a$) in the presence of external magnetic field can be expressed  {by the differential equation}  \cite{Raffelt:1987im, raffelt, raffeltbook}
\begin{align}
\left(\omega+\left(
\begin{array}{ccc}
\De & 0 & \Dagax\\
0 & \De & \Dagay\\
\Dagax & \Dagay & \Da\\
\end{array}
\right)+i\partial_z\right)
\left(
\begin{array}{c}
A_{x}\\
A_{y}\\
a\\
\end{array}
\right)=0,\label{eq-evolution}
\end{align}
where, we have assumed that the photons are propagating in the z-direction and, 
\begin{align}
\begin{split}
\left(\frac{\Delta^{ x,y}_{\gammaa}}{{\rm{Mpc^{-1}}}}\right) &\equiv \frac{g_{\gammaa} |B_{x,y}|}{2} = 15.2
\left(\frac{g_{\gammaa}}{10^{-11}{\rm Gev}^{-1}}\right)\left(\frac{B_{x,y}}{\mu{\rm G}}\right),\\
\left(\frac{{\Da }}{{\rm{Mpc^{-1}}}}\right) &\equiv - \frac{m^2_a}{2\omega}= -1.9\times
10^4\left(\frac{\ma}{10^{-14}{\rm eV}}\right)\left(\frac{100~ {\rm
      GHz}}{\nu}\right),\\
        \left(\frac{\De}{{\rm{Mpc^{-1}}}}\right) &\approx   \frac{\op^2}{2\omega}
\left[-1 +7.3\times 10^{-3}\frac{\NH}{\Ne}  \left(\frac{\omega}{{\rm
        eV}}\right)^2\right] \\
&=-2.6\times
10^6 \left(\frac{\Ne}{10^{-5}{\rm cm^{-3}}}\right) \left(\frac{100~{\rm
    GHz}}{\nu}\right), 
\end{split}
\end{align}
where   {$g_{\gammaa}$ is the photon-ALP coupling strength}, $\op=[4\pi \alpha\Ne/(\me)]^{1/2}$ is the plasma frequency, $m_e$ is the mass of electron, $n_e$ is the electron density, $m_a$ is the mass of ALPs, $B_{x,y}$ is the magnetic field along the direction $x$ or $y$ and $n_H$ is the density of the hydrogen atoms.  We neglect  Faraday rotation. For homogeneous magnetic field and electron density, the above equation reduces to the simple form \cite{Raffelt:1987im, raffelt, raffeltbook}
\begin{align}\label{prob-pa}
P(\gamma \rightarrow a) & = \frac{(\Delta_{\gammaa}s)^2}{(\Dosc s/2)^2}
\sin^2(\Dosc s/2), 
\end{align}
where $\Dosc^2 = (\Da  - \De )^2 +4\Delta^2_{\gammaa}$ and $s$ is the length along the line of sight.  {The mixing angle $\theta$ can be defined as $\sin(2\theta)= 2\Delta_{\gammaa}/\Dosc$}. For the case of inhomogeneous magnetic field and electron density, we need to solve Eq. \eqref{eq-evolution} along the line of sight, to calculate the probability of conversion. 

It is possible to obtain approximate analytical solutions for Milky Way, galaxy clusters and voids in the limit of stochastic magnetic fields with electron density changing slowly compared to the magnetic fields  {such that the adiabiticity parameter $\gamma_{ad} \equiv |\frac{\Dosc}{2\bigtriangledown\theta}|<1$ \cite{Mukherjee:2018oeb}}
 {\begin{align}
\begin{split}
{P}(\gamma \rightarrow a)
&\approx \frac{\Daga^2 R d }{2} = 10^{-5}\left(\frac{g_{\gammaa}}{10^{-9}~{\rm
      GeV}^{-1}}\right)^2\left(\frac{B_T}{1~{\rm
      \mu G}}\right)^2\left(\frac{R}{~{\rm 10kpc}}\right)\left(\frac{d}{10^{-3}~{\rm pc}}\right) \text{for Milky Way,}\label{Eq:ismlimit}\\
      \bar{P}(\gamma \rightarrow a)&\approx \frac{\Daga^2\Rv d }{2}= 10^{-6}\left(\frac{g_{\gammaa}}{10^{-10}~{\rm
      GeV}^{-1}}\right)^2\left(\frac{B_T}{1~{\rm
      nG}}\right)^2\left(\frac{\Rv}{10~{\rm Mpc}}\right)\left(\frac{d}{10~{\rm pc}}\right)\text{for voids,}\\
      \bar{P}(\gamma \rightarrow a)&\approx \frac{\Daga^2\Rc d }{2}= 10^{-5}\left(\frac{g_{\gammaa}}{10^{-9}~{\rm
      GeV}^{-1}}\right)^2\left(\frac{B_T}{0.1~{\rm
      \mu G}}\right)^2\left(\frac{\Rc}{10~{\rm Mpc}}\right)\left(\frac{d}{10^{-4}~{\rm pc}}\right)\text{for galaxy clusters,}
     \end{split}
\end{align}}
where $B_T$ is the magnetic field transverse to the line of sight in the domain of size $d$ and $R$ is the typical size of the region being considered ($\Rv$ and $\Rc$ corresponds to the typical size considered for voids and galaxy clusters).  {In the above equation we have used the value of magnetic field which are consistent with the recent observations from synchrotron emission map and rotation measures \cite{2012ApJ...757...14J, 2012ApJ...761L..11J} and for galaxy clusters, the value of magnetic field are inferred from the Faraday rotation \cite{Boehringer:2016kqe}}. 

The above equation shows that the photon-axion coupling strength $g_{\gammaa}$ is degenerate with  astrophysical parameters like $B_T, R$ and $d$. The change in the intensity of the CMB is given by 
 {
\begin{align}\label{spectrum}
\begin{split}
\Delta I^{\gammaa}({\nu}, \hat p)=&  {P}(\gamma \rightarrow a, \hat p)\left(\frac{2h\nu^3}{c^2}\right)\frac{1}{(e^x-1)},\\
=&  \alpha(\hat p) \left(\frac{2h\nu^3}{c^2}\right)\frac{1}{(e^x-1)},
\end{split}
\end{align}}
 {where, $x= h\nu/(k_BT_{CMB})$, $T_{CMB}= 2.7255$ K and $h$, $c$, and $k_B$ are Planck's constant, speed of light and Boltzmann constant respectively. In Eq. \eqref{spectrum}, ${P}(\gamma \rightarrow a, \hat p) \equiv \alpha (\hat p)$ is the amplitude of the distortion  {along the direction denoted by $\hat p$}.  {The direction dependence of  ${P}(\gamma \rightarrow a, \hat p)$ in the above equation arises due to the direction dependence of the magnetic field strength.} All our results will be for this amplitude which varies over the sky while the shape of the distortion is fixed.} 

The spatially fluctuating spectral distortions of the CMB can be measured by experiments without an absolute calibrator but having multiple frequency channels, such as WMAP \cite{2013ApJS..208...20B} and Planck \cite{2016A&A...594A...1P}.  
 
\section{{Component separation for the axion spectrum}}
The Planck satellite measured the differential sky intensity in nine frequency channels covering the frequency range $30$-$857$ GHz. The sky signal is a combination of several components including Galactic foregrounds (like synchrotron, free-free, AME, galactic dust), CMB, thermal Sunyaev-Zeldovich (tSZ) \cite{zeldovich,sz1970}, and Cosmic Infrared Background (CIB). A number of algorithms have been developed over the past decades to separate the observed sky signal into different components \cite{Eriksen:2005dr,Delabrouille:2008qd, Rogers:2016rdb, Delabrouille:2002kz, Vansyngel:2014dfa, Tegmark:1995pn,Eriksen:2004jg,Hurier:2010wu, 2011MNRAS.410.2481R,Khatri:2014sra,Khatri:2018fjk}. 

 In this analysis, we consider six frequency channels ($70, 100, 143, 217, 353, 545$ GHz) to obtain the sky-map for the  {ASD} signal.  {Channels below $70$ GHz and above $545$ GHz are highly contaminated by synchrotron emission/AME, and dust respectively.} So, we only consider these six channels in this analysis. These six frequency channels are also not completely clean and are dominated by foreground contaminations in the galactic plane. There are also point source contaminations of both galactic and extragalactic origin. We will use the ILC algorithm to separate the axion distortion from other cosmological and Galactic components  \cite{2003ApJS..148....1B, Eriksen:2004jg}. In order to remove the worst Galactic and point source contamination, we apply a mask on the full sky map and use only partial sky in the analysis. We consider two different masks  having usable sky-fraction $f_{sky}= 27\%$ and $45\%$ \cite{Khatri:2015tla}, created specially to search for new spectral distortions in the Planck data. The masks are publicly available \cite{mask}. The 45\%  mask is shown in Fig.~\ref{Fig:maphs}.
 
\begin{figure}
\centering
     \includegraphics[trim={0cm 0cm 0cm 0.cm}, clip, width=1.\linewidth]{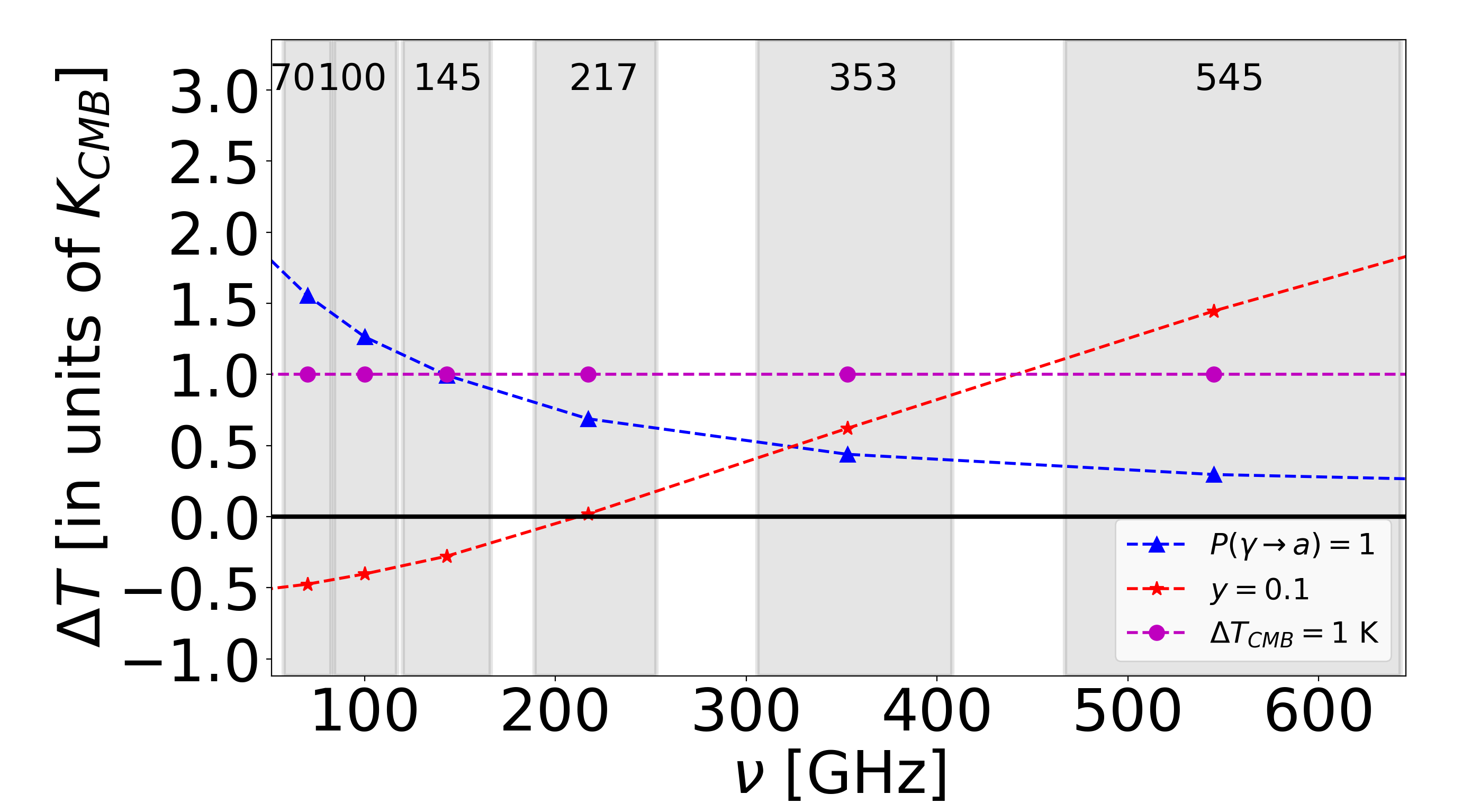}
     \captionsetup{singlelinecheck=on,justification=raggedright}
 \caption{ {Frequency spectrum of the non-resonance photon-axion signal (in blue), $y$-type distortion for $y=0.1$ (in red) and CMB fluctuation for $\Delta T= 1 K_{CMB}$ (
 in magenta) are plotted in thermodynamic temperature units ($K_{CMB}$) using Eq. \eqref{ilc-2}. The grey bands indicate the frequency channels used in this analysis and the central frequency is mentioned in the top.}}\label{Fig:spec}
\end{figure}

 These masks remove the point source contaminations (tSZ, CO line emission) along with the most contaminated region of the Galaxy. 
The Planck frequency sky maps are in the CMB temperature units ($K_{CMB}$) except for $545$ GHz map which has units of MJy/Sr and which we also convert into $K_{CMB}$ units \cite{Adam:2015vua}. The different frequency channels also have a finite transmission bandwidth ($w_{\nu} (\nu')$) \cite{pla}.  So in order to extract the signal with a particular spectrum, we need to convert the spectrum from intensity to $K_{\text{CMB}}$ units  in the particular frequency band by integrating over the transmission function, using the relation \cite{2014A&A...571A...9P}
\begin{equation}\label{ilc-2}
\Delta T^{\beta} (\nu) = \frac{\int w_{\nu}(\nu') I^{\beta}({\nu'}) d\nu'}{\int w_{\nu}(\nu')I'^{pl}({\nu'}) d\nu'}\,\, \text{[in units of $K_{CMB}$]},
\end{equation}
where, $I'^{pl}({\nu}) \equiv  \frac{\partial I^{pl}({\nu})}{\partial T}$ and $\beta \in$ \{CMB,  {ASD}, y-type distortions (SZ), $\mu$-type\}. 
The bandpass corrected frequency spectrum of the  {ASD}, CMB and SZ are shown in Fig.~\ref{Fig:spec} along with the frequency bands used in the analysis. 

The observed sky signal at different frequencies ($S_{\nu_i}$) can be modeled in terms of the multiple components as
\begin{align}\label{ilc-1}
\begin{split}
\mathbf{S} (\hat p)&= \mathbf{A}\mathbf{x}(\hat p) + \mathbf{n} (\hat p),
\end{split}
\end{align}
here $\mathbf{A}$ is the mixing matrix $[\mathbf{a_1, a_2, \hdots, a_{M}}]$ with dimension $N \times M$, where $N$ is the number of frequency channels and $M$ is the number of components, $\mathbf{a_i}$ is the spectrum of the $i^{th}$ component, $\mathbf{n}$ is the noise at pixel $\hat p$. 
The ILC solution for the axion signal with known spectrum $\mathbf{a_{\gamma a}}$ is given by the linear combination of the input maps  \cite{2003ApJS..148....1B, Eriksen:2004jg}
\begin{equation}\label{ilc-1a}
\alpha =  \mathbf{\hat W}_{\gamma a}^T \mathbf{S}(\hat p).    
\end{equation}
where, $ \mathbf{\hat W}_{\gamma a}(\nu)=  \mathbf{C}_S^{-1} \mathbf{a}_{\gamma a} (\mathbf{a}^T_{\gamma a} \mathbf{C}_S^{-1}\mathbf{a}_{\gamma a})^{-1}$ and $\mathbf{C}_S = \langle \mathbf{S} \mathbf{S}^T\rangle$\footnote{ The angular bracket denotes average over pixels.} is the covariance matrix of the data inferred from the masked sky maps. We have subtracted the global mean of unmasked pixels from each map before performing ILC i.e $\langle S\rangle=0$. 

\section{{Results}}
Applying the above mentioned component separation method to half ring \footnote{ {Half ring maps are the Planck maps produced from the first (or second) half of the pointing period.}} maps, we obtain the half ring ALP distortion maps  using $70-545$ GHz sky maps of Planck, all smoothed to a common angular resolutions ($15$ or $20$ arcmin) and combine them to get the  
half-ring-half-sum (HRHS) and half-ring-half-difference (HRHD) maps. The HRHS map includes both signal and noise, whereas the HRHD gives the noise estimate in the HRHS map. 

\begin{figure}
\centering
\includegraphics[trim={2cm 0cm 1cm 3cm}, clip, width=1.\linewidth]{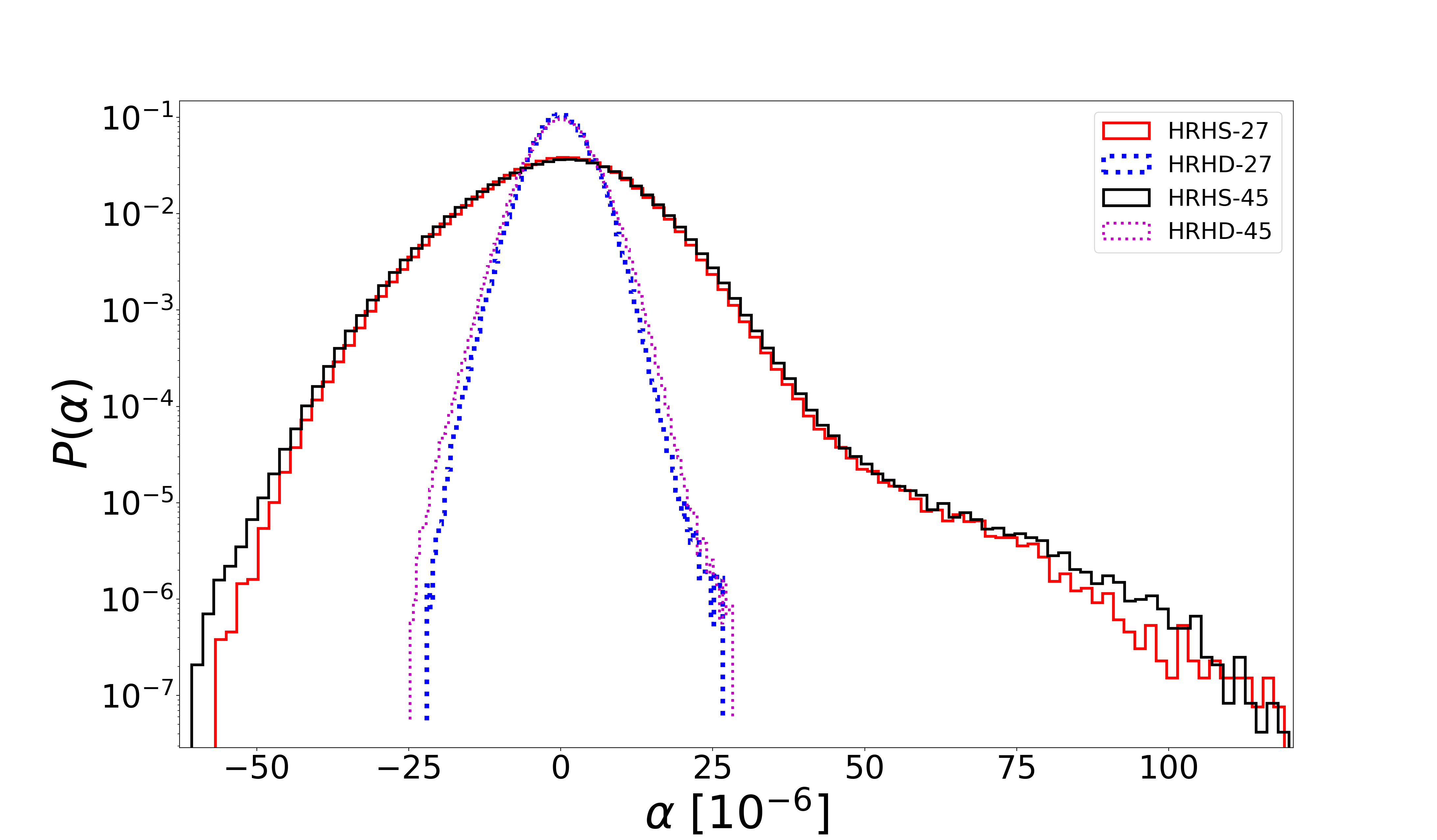}
     \captionsetup{singlelinecheck=on,justification=raggedright}
 \caption{The 1-D probability distribution function of the masked HRHS and HRHD maps are depicted for two $f_{sky}$ values. The HRHS maps contain contamination from other components. The HRHD maps contain only noise and are close to the expected Gaussian distribution. }\label{Fig:hist}
\end{figure}

The HRHS map for sky fraction $45\%$ is shown in Fig.~\ref{Fig:maphs}. We plot the 1-D Probability Distribution Function (PDF) in Fig.~\ref{Fig:hist}.  There are a significant number of pixels above the Gaussian HRHD noise PDF making the HRHS PDF broader with a significant positive tail. All (or most) of the signal is contamination from other components such as CMB, SZ, dust as well as unresolved point sources.  {As a result this map is only an upper bound on $\alpha$-distortion}. For $f_{sky}=27\%$ and $f_{sky}=45\%$ with $15$ arcmin smoothing scale, the $95^{th}$ percentile upper limits from HRHS maps are $17.3 \times 10^{-6}$ and $18.5 \times 10^{-6}$ respectively. These bounds are conservative upper limits on the ASD signal which include contaminations from instrument noise as well as astrophysical and cosmological signals. 

Assuming that the signal is dominated by contamination from other components, we can put an upper limit to the RMS ALPs distortions, $\sigma_{RMS}^{\alpha}$, after removing the noise contribution, $\sigma_{RMS}^{\alpha}= (\sigma_{HRHS}^2 - \sigma_{HRHD}^2)^{1/2}$. The upper limits on the ALPs distortion for different resolutions is shown in Table \ref{tab:1}. 

\begin{table}[h]
\centering
\captionof{table}{$\sigma_{RMS}^{\alpha}$ in units of $10^{-6}$} \label{tab:1}
\begin{tabular}{ |r|m{2.3cm}|m{2.2cm}|m{2.2cm}|}
\hline
\centering
$f_{sky}$&smoothing scale (in arcmin) & 70-545 (GHz) $\sigma_{RMS}^{\alpha}$ & 100-545 (GHz) $\sigma_{RMS}^{\alpha}$\\
\hline
 \centering
0.27&15  &  $10.3$ & $10.7$ \\\cline{2-4}
&20 &  $7.9 $ &  $8.1$\\ 
\hline
  \centering
0.45&15  &   $10.6$ & $11.1$ \\ \cline{2-4}
&20 & $8.1 $  & $8.3$\\ 
\hline
\end{tabular}
\end{table}

These constraints are obtained by only using the frequency spectrum of the  {ASD} signal and without assuming any model of electron density and magnetic field. 
However, in order to convert these constraints into constraints on photon-axion coupling strength $g_{\gamma a}$, we need  a model of the turbulent electron density and magnetic field of our Galaxy. A further complication is that these constraints are for the fluctuations of the  {ASD} (RMS) i.e. fluctuation of probability of conversion defined in Eq. \eqref{Eq:ismlimit} and not the average  {ASD}.  {With the assumption that the fluctuations in the signal are of the same order as its average value \cite{2012ApJ...757...14J}, we can translate the $1-\sigma^\alpha_{RMS}$ bound on ASD into a bound on $g_{\gammaa}$. For a typical $\sigma^\alpha_{RMS}\lesssim 10^{-5}$ translates into a bound on photon axion coupling of $g_{\gammaa} \lesssim 10^{-9}~{\rm GeV}^{-1}$ for Milky Way using Eq. \eqref{Eq:ismlimit}. The $1-\sigma^\alpha_{RMS}$ bound obtained from the CMB maps on $P(\gamma \rightarrow a)$ does not alter with the change in the magnetic field model and depends only on the spectral shape of the ASD signal. However, in order to relate the constraints on $P(\gamma \rightarrow a)$ with $g_{\gammaa}$, we require a model of the magnetic field. The measurements of the galactic magnetic field are made with about $25\%-30\%$ error-bars \cite{2012ApJ...757...14J,2012ApJ...761L..11J}. So, our bound of $g_{\gammaa}$ can vary within the uncertainty of the magnetic field.} 
The current particle physics bounds from the CERN ALP Solar Telescope (CAST) is $g_{\gammaa} < 6.6 \times 10^{-11}$ GeV$^{-1}$ at $95\%$ C.L. \cite{Anastassopoulos:2017ftl}. The bound obtained from the Planck data provides an independent but a weaker bound than the current bound from CAST.

\begin{figure}[h]
\centering
\begin{subfigure}{1.\linewidth}
\includegraphics[trim={0cm 0cm 1cm 0.8cm}, clip, width=0.9\textwidth]{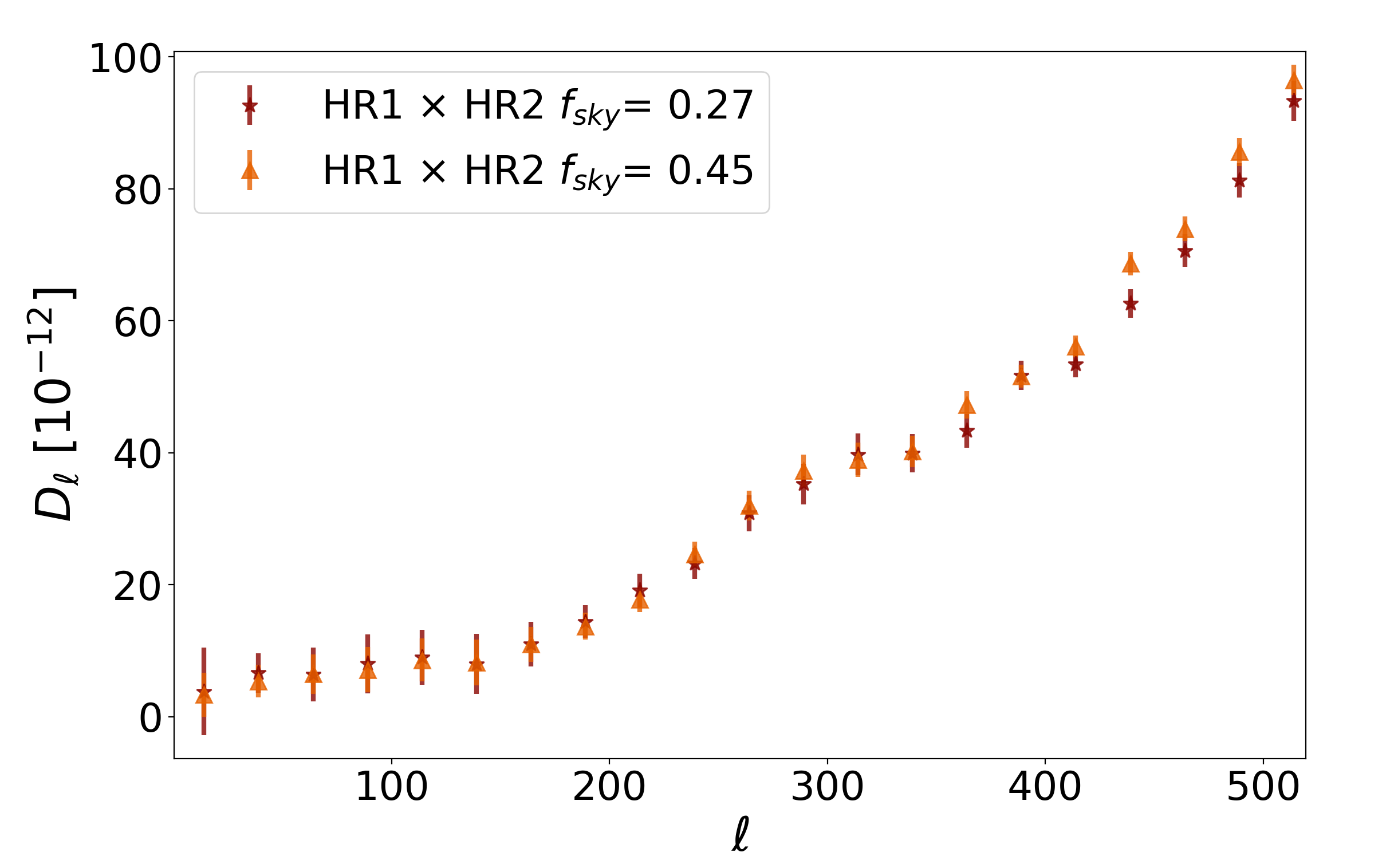}
     \caption{}
     \end{subfigure}
     \captionsetup{singlelinecheck=on,justification=raggedright}
 \caption{The angular power spectrum $D_l\equiv l(l+1)\hat C_l/2\pi$ for $HR1 \times HR2$ for two different masked sky mask ($f_{sky}= 0.27$ and $0.45$).}\label{Fig:powersp}
\end{figure}

We can also calculate the angular power spectrum of the  {ASD} map providing upper bounds on ALPs distortion fluctuations on different angular scales. We calculate the cross-power spectrum of the half-ring maps using PolSpice \cite{2004MNRAS.349..603E, polspice} with the mask apodised by a $30$ arcmin Gaussian \cite{Khatri:2015tla}. The power spectrum $D_l= l(l+1)\hat C_l/2\pi$, where $\hat C_l= \sum_{m} \alpha^{HR1}_{lm}\alpha^{*HR2}_{lm}/(2l+1)$ 
and, $\alpha_{lm}$ is the spherical harmonic transform of the ALP distortion map,  is shown (after correction for the effects of mask and beam\cite{Hivon:2001jp,2001ApJ...548L.115S, 2004MNRAS.350..914C, polspice})  in Fig.~\ref{Fig:powersp}. The Gaussian error-bars on $D_l$ are the analytical estimates obtained using PolSpice \cite{2004MNRAS.349..603E, polspice}. 

\section{{Conclusion}}
In this paper, we provide the first observational constraints on the non-resonant photon-ALP conversion (or Axion Spectral Distortion ( {ASD})) using Planck data. The  {ASD} can be created in the Milky Way when the CMB photons travel through the turbulent magnetic field in the galactic halo and get converted to light spin-0 particles such as light axion particles ($m_a << E_{CMB}$) or light scalars. 
Since both the stochastic magnetic field and the electron density have large fluctuations, the induced spectral distortions will vary on the sky creating a spatially fluctuating unpolarized spectral distortions. The unpolarized  {ASD} has a unique spectral shape different from CMB and other known spectral distortions signal such as $y$-type distortions. 

Using six frequency channels ($70, 100, 143, 217, 353, 545$) GHz of the Planck satellite, we obtain the sky-map for the  {ASD} using the ILC algorithm. In order to minimize the contaminations, 
 we mask the most contaminated sky with two different masks having the unmasked sky fractions $27\%$ and $45\%$. The  sky map of the  {ASD} is shown in Fig.~\ref{Fig:maphs}. These maps are dominated by the residual contaminations from other components. Hence, we can only provide upper limits on the  {ASD}  shown in Table~\ref{tab:1}. These are robust constraints on the fluctuation of probability of conversion from photon to axion. We expect the fluctuations of  {ASD} on the sky to be of the order unity since the stochastic magnetic fields as well as the electron density have large fluctuations in our own Galaxy as well as outside it where we expect larger contributions from the directions of the nearby voids and smaller distortions from other directions. We can therefore assume that average distortions from our own Galaxy and nearby voids is of the same order of magnitude as the RMS fluctuations. Under this assumption the upper limit of $10.6\, \times 10^{-6}$ can be translated into combined limits on $g_{\gamma a}B_T$ using Eq. \eqref{Eq:ismlimit}.  
  A future data-driven model of the galactic magnetic field in future will  allow  making more precise statements. 
Future experiments such as Simons Observatory \cite{2018arXiv180804493G}, Simons Array \cite{2016JLTP..184..805S}, Adv-ACT \cite{2016ApJS..227...21T}, SPT-3G \cite{2016JLTP..184..805S} and  proposed missions like CMB-S4 \cite{Abazajian:2016yjj}, LiteBIRD \cite{Matsumura:2016sri}, CMB-Bharat, PIXIE \cite{2014AAS...22343901K} and PICO \cite{2018arXiv180801369Y} 
will improve    these constraints significantly. 
Future CMB experiments having more frequency channels will play a crutial role in removing the contamination from CMB fluctuations and the galactic foregrounds. For Planck number of frequency channels is the main limitation.  In particular we are limited by foreground contamination and not by the detector noise. 
Lower detector noise and higher angular resolution than Planck will also play an important role to reduce the noise in the recovered ASD maps and improving the constraints on $g_{\gamma a}$.
The polarization data of  Planck (and also of ground based experiments), is capable of  imposing constraints on the resonant photon-axion conversion \cite{Mukherjee:2018oeb}, which can directly constrain the photon-axion coupling $g_{\gamma a}$ given a model of the Galactic magnetic field. We will address the polarized  {ASD} in a future analysis by using the Planck-2018 polarization data.

\textbf{Acknowledgement}
SM would like to thank  Nick Battaglia, Siguard Naess, Joseph Silk and David Spergel for useful discussions. The Flatiron Institute is supported by the Simons Foundation. The work of SM and BDW is also supported by the Labex ILP (reference ANR-10-LABX-63) part of the Idex SUPER, and received financial state aid managed by the Agence Nationale de la Recherche, as part of the programme Investissements d'avenir under the reference ANR-11-IDEX-0004-02. The work of BDW is also supported by the grant ANR-16-CE23-0002.  RK was supported by SERB grant no. ECR/2015/000078 from Science and Engineering Research Board, Department of Science and Technology, Govt. of India and MPG-DST partner group between Max Planck Institute for Astrophysics, Garching and Tata Institute for Fundamental Research, Mumbai funded by Max Planck Gesellschaft.
\bibliography{planck-axion}

\providecommand{\href}[2]{#2}\begingroup\raggedright\begin{thebibliography}{10}

\bibitem{zeldovich}
Y.~B. {Zeldovich} and R.~A. {Sunyaev}, \emph{{The Interaction of Matter and
  Radiation in a Hot-Model Universe}},
  \href{https://doi.org/10.1007/BF00661821}{\emph{\apss} {\bfseries 4} (1969)
  301}.

\bibitem{sz1970}
R.~A. {Sunyaev} and Y.~B. {Zeldovich}, \emph{{The interaction of matter and
  radiation in the hot model of the Universe, II}},
  \href{https://doi.org/10.1007/BF00653472}{\emph{\apss} {\bfseries 7} (1970)
  20}.

\bibitem{Chluba:2011hw}
J.~{Chluba} and R.~A. {Sunyaev}, \emph{{The evolution of CMB spectral
  distortions in the early Universe}},
  \href{https://doi.org/10.1111/j.1365-2966.2011.19786.x}{\emph{\mnras}
  {\bfseries 419} (2012) 1294}.

\bibitem{Chluba:2012gq}
J.~{Chluba}, R.~{Khatri} and R.~A. {Sunyaev}, \emph{{CMB at 2 {$\times$} 2
  order: the dissipation of primordial acoustic waves and the observable part
  of the associated energy release}},
  \href{https://doi.org/10.1111/j.1365-2966.2012.21474.x}{\emph{\mnras}
  {\bfseries 425} (2012) 1129}
  [\href{https://arxiv.org/abs/1202.0057}{{\ttfamily 1202.0057}}].

\bibitem{Khatri:2012tv}
R.~{Khatri} and R.~A. {Sunyaev}, \emph{{Creation of the CMB spectrum: precise
  analytic solutions for the blackbody photosphere}},
  \href{https://doi.org/10.1088/1475-7516/2012/06/038}{\emph{\jcap} {\bfseries
  6} (2012) 038} [\href{https://arxiv.org/abs/1203.2601}{{\ttfamily
  1203.2601}}].

\bibitem{Khatri:2012rt}
R.~{Khatri}, R.~A. {Sunyaev} and J.~{Chluba}, \emph{{Mixing of blackbodies:
  entropy production and dissipation of sound waves in the early Universe}},
  \href{https://doi.org/10.1051/0004-6361/201219590}{\emph{\aap} {\bfseries
  543} (2012) A136} [\href{https://arxiv.org/abs/1205.2871}{{\ttfamily
  1205.2871}}].

\bibitem{Khatri:2012tw}
R.~{Khatri} and R.~A. {Sunyaev}, \emph{{Beyond y and {$\mu$}: the shape of the
  CMB spectral distortions in the intermediate epoch, $1.5 {\times} 10^{4}
  \lesssim z \lesssim 2 {\times} 10^{5}$}},
  \href{https://doi.org/10.1088/1475-7516/2012/09/016}{\emph{\jcap} {\bfseries
  9} (2012) 016} [\href{https://arxiv.org/abs/1207.6654}{{\ttfamily
  1207.6654}}].

\bibitem{sk2013}
R.~A. {Sunyaev} and R.~{Khatri}, \emph{{Unavoidable CMB Spectral Features and
  Blackbody Photosphere of Our Universe}},
  \href{https://doi.org/10.1142/S0218271813300140}{\emph{International Journal
  of Modern Physics D} {\bfseries 22} (2013) 1330014}
  [\href{https://arxiv.org/abs/1302.6553}{{\ttfamily 1302.6553}}].

\bibitem{Chluba:2016bvg}
J.~{Chluba}, \emph{{Which spectral distortions does {$\Lambda$}CDM actually
  predict?}}, \href{https://doi.org/10.1093/mnras/stw945}{\emph{\mnras}
  {\bfseries 460} (2016) 227}
  [\href{https://arxiv.org/abs/1603.02496}{{\ttfamily 1603.02496}}].

\bibitem{Hill:2015tqa}
J.~C. {Hill}, N.~{Battaglia}, J.~{Chluba}, S.~{Ferraro}, E.~{Schaan} and D.~N.
  {Spergel}, \emph{{Taking the Universe's Temperature with Spectral Distortions
  of the Cosmic Microwave Background}},
  \href{https://doi.org/10.1103/PhysRevLett.115.261301}{\emph{Physical Review
  Letters} {\bfseries 115} (2015) 261301}
  [\href{https://arxiv.org/abs/1507.01583}{{\ttfamily 1507.01583}}].

\bibitem{Emami:2015xqa}
R.~{Emami}, E.~{Dimastrogiovanni}, J.~{Chluba} and M.~{Kamionkowski},
  \emph{{Probing the scale dependence of non-Gaussianity with spectral
  distortions of the cosmic microwave background}},
  \href{https://doi.org/10.1103/PhysRevD.91.123531}{\emph{\prd} {\bfseries 91}
  (2015) 123531} [\href{https://arxiv.org/abs/1504.00675}{{\ttfamily
  1504.00675}}].

\bibitem{Hasselfield:2013wf}
M.~Hasselfield et~al., \emph{{The Atacama Cosmology Telescope:
  Sunyaev-Zel'dovich selected galaxyclusters at 148 GHz from three seasons of
  data}}, \href{https://doi.org/10.1088/1475-7516/2013/07/008}{\emph{JCAP}
  {\bfseries 1307} (2013) 008}
  [\href{https://arxiv.org/abs/1301.0816}{{\ttfamily 1301.0816}}].

\bibitem{Bleem:2014iim}
{\scshape SPT} collaboration, \emph{{Galaxy Clusters Discovered via the
  Sunyaev-Zel'dovich Effect in the 2500-square-degree SPT-SZ survey}},
  \href{https://doi.org/10.1088/0067-0049/216/2/27}{\emph{Astrophys. J. Suppl.}
  {\bfseries 216} (2015) 27} [\href{https://arxiv.org/abs/1409.0850}{{\ttfamily
  1409.0850}}].

\bibitem{Ade:2015gva}
{\scshape Planck} collaboration, \emph{{Planck 2015 results. XXVII. The Second
  Planck Catalogue of Sunyaev-Zeldovich Sources}},
  \href{https://doi.org/10.1051/0004-6361/201525823}{\emph{Astron. Astrophys.}
  {\bfseries 594} (2016) A27}
  [\href{https://arxiv.org/abs/1502.01598}{{\ttfamily 1502.01598}}].

\bibitem{Staniszewski:2008ma}
Z.~Staniszewski et~al., \emph{{Galaxy clusters discovered with a
  Sunyaev-Zel'dovich effect survey}},
  \href{https://doi.org/10.1088/0004-637X/701/1/32}{\emph{Astrophys. J.}
  {\bfseries 701} (2009) 32} [\href{https://arxiv.org/abs/0810.1578}{{\ttfamily
  0810.1578}}].

\bibitem{2010A&A...518L..16Z}
{M. {Zemcov} et al.}, \emph{{First detection of the Sunyaev Zel'dovich effect
  increment at {$\lambda$} $<$ 650 {$\mu$}m}},
  \href{https://doi.org/10.1051/0004-6361/201014685}{\emph{\aap} {\bfseries
  518} (2010) L16} [\href{https://arxiv.org/abs/1005.3824}{{\ttfamily
  1005.3824}}].

\bibitem{2013A&A...550A.134P}
{Planck Collaboration}, P.~A.~R. {Ade}, N.~{Aghanim}, M.~{Arnaud},
  M.~{Ashdown}, F.~{Atrio-Barandela} et~al., \emph{{Planck intermediate
  results. VIII. Filaments between interacting clusters}},
  \href{https://doi.org/10.1051/0004-6361/201220194}{\emph{\aap} {\bfseries
  550} (2013) A134} [\href{https://arxiv.org/abs/1208.5911}{{\ttfamily
  1208.5911}}].

\bibitem{Mukherjee:2018oeb}
S.~Mukherjee, R.~Khatri and B.~D. Wandelt, \emph{{Polarized anisotropic
  spectral distortions of the CMB: Galactic and extragalactic constraints on
  photon-axion conversion}},
  \href{https://doi.org/10.1088/1475-7516/2018/04/045}{\emph{JCAP} {\bfseries
  1804} (2018) 045} [\href{https://arxiv.org/abs/1801.09701}{{\ttfamily
  1801.09701}}].

\bibitem{firas}
D.~J. {Fixsen}, E.~S. {Cheng}, J.~M. {Gales}, J.~C. {Mather}, R.~A. {Shafer}
  and E.~L. {Wright}, \emph{{The Cosmic Microwave Background Spectrum from the
  Full COBE FIRAS Data Set}}, \href{https://doi.org/10.1086/178173}{\emph{\apj}
  {\bfseries 473} (1996) 576}.

\bibitem{firasan}
D.~J. {Fixsen}, G.~{Hinshaw}, C.~L. {Bennett} and J.~C. {Mather}, \emph{{The
  Spectrum of the Cosmic Microwave Background Anisotropy from the Combined COBE
  FIRAS and DMR Observations}},
  \href{https://doi.org/10.1086/304560}{\emph{\apj} {\bfseries 486} (1997) 623}
  [\href{https://arxiv.org/abs/astro-ph/9704176}{{\ttfamily
  astro-ph/9704176}}].

\bibitem{fm2002}
D.~J. {Fixsen} and J.~C. {Mather}, \emph{{The Spectral Results of the
  Far-Infrared Absolute Spectrophotometer Instrument on COBE}},
  \href{https://doi.org/10.1086/344402}{\emph{\apj} {\bfseries 581} (2002)
  817}.

\bibitem{2009ApJ...707..916F}
D.~J. {Fixsen}, \emph{{The Temperature of the Cosmic Microwave Background}},
  \href{https://doi.org/10.1088/0004-637X/707/2/916}{\emph{\apj} {\bfseries
  707} (2009) 916} [\href{https://arxiv.org/abs/0911.1955}{{\ttfamily
  0911.1955}}].

\bibitem{Khatri:2015jxa}
R.~{Khatri} and R.~{Sunyaev}, \emph{{Limits on the fluctuating part of y-type
  distortion monopole from Planck and SPT results}},
  \href{https://doi.org/10.1088/1475-7516/2015/08/013}{\emph{\jcap} {\bfseries
  8} (2015) 013} [\href{https://arxiv.org/abs/1505.00781}{{\ttfamily
  1505.00781}}].

\bibitem{Pajer:2013oca}
E.~Pajer and M.~Zaldarriaga, \emph{{A hydrodynamical approach to CMB
  $\mu$-distortion from primordial perturbations}},
  \href{https://doi.org/10.1088/1475-7516/2013/02/036}{\emph{JCAP} {\bfseries
  1302} (2013) 036} [\href{https://arxiv.org/abs/1206.4479}{{\ttfamily
  1206.4479}}].

\bibitem{Khatri:2015tla}
R.~{Khatri} and R.~{Sunyaev}, \emph{{Constraints on {$\mu$}-distortion
  fluctuations and primordial non-Gaussianity from Planck data}},
  \href{https://doi.org/10.1088/1475-7516/2015/09/026}{\emph{\jcap} {\bfseries
  9} (2015) 026} [\href{https://arxiv.org/abs/1507.05615}{{\ttfamily
  1507.05615}}].

\bibitem{Arvanitaki:2009fg}
A.~Arvanitaki, S.~Dimopoulos, S.~Dubovsky, N.~Kaloper and J.~March-Russell,
  \emph{{String Axiverse}},
  \href{https://doi.org/10.1103/PhysRevD.81.123530}{\emph{Phys. Rev.}
  {\bfseries D81} (2010) 123530}
  [\href{https://arxiv.org/abs/0905.4720}{{\ttfamily 0905.4720}}].

\bibitem{Anastassopoulos:2017ftl}
{\scshape CAST} collaboration, \emph{{New CAST Limit on the Axion-Photon
  Interaction}}, \href{https://doi.org/10.1038/nphys4109}{\emph{Nature Phys.}
  {\bfseries 13} (2017) 584}
  [\href{https://arxiv.org/abs/1705.02290}{{\ttfamily 1705.02290}}].

\bibitem{Bastidon:2015efa}
{\scshape ALPS II} collaboration, \emph{{Any Light Particle Search II - Status
  Overview}},  in \emph{{Proceedings, 11th Patras Workshop on Axions, WIMPs and
  WISPs (Axion-WIMP 2015): Zaragoza, Spain, June 22-26, 2015}}, pp.~31--34,
  2015, \href{https://arxiv.org/abs/1509.02070}{{\ttfamily 1509.02070}},
  \href{https://doi.org/10.3204/DESY-PROC-2015-02/bastidon_noemie_talk}{DOI}.

\bibitem{Majorovits:2016yvk}
{\scshape MADMAX Working Group} collaboration, \emph{{MADMAX: A new Dark Matter
  Axion Search using a Dielectric Haloscope}},  in \emph{{Proceedings, 12th
  Patras Workshop on Axions, WIMPs and WISPs (PATRAS 2016): Jeju Island, South
  Korea, June 20-24, 2016}}, pp.~94--97, 2017,
  \href{https://arxiv.org/abs/1611.04549}{{\ttfamily 1611.04549}},
  \href{https://doi.org/10.3204/DESY-PROC-2009-03/Majorovits_Bela}{DOI}.

\bibitem{2010PhRvL.104d1301A}
S.~J. {Asztalos}, G.~{Carosi}, C.~{Hagmann}, D.~{Kinion}, K.~{van Bibber},
  M.~{Hotz} et~al., \emph{{SQUID-Based Microwave Cavity Search for Dark-Matter
  Axions}}, \href{https://doi.org/10.1103/PhysRevLett.104.041301}{\emph{\prl}
  {\bfseries 104} (2010) 041301}
  [\href{https://arxiv.org/abs/0910.5914}{{\ttfamily 0910.5914}}].

\bibitem{2014PhRvX...4b1030B}
D.~{Budker}, P.~W. {Graham}, M.~{Ledbetter}, S.~{Rajendran} and A.~O.
  {Sushkov}, \emph{{Proposal for a Cosmic Axion Spin Precession Experiment
  (CASPEr)}}, \href{https://doi.org/10.1103/PhysRevX.4.021030}{\emph{Physical
  Review X} {\bfseries 4} (2014) 021030}
  [\href{https://arxiv.org/abs/1306.6089}{{\ttfamily 1306.6089}}].

\bibitem{Tegmark:1995pn}
M.~Tegmark and G.~Efstathiou, \emph{{A method for subtracting foregrounds from
  multi-frequency cmb sky maps}},
  \href{https://doi.org/10.1093/mnras/281.4.1297}{\emph{Mon. Not. Roy. Astron.
  Soc.} {\bfseries 281} (1996) 1297}
  [\href{https://arxiv.org/abs/astro-ph/9507009}{{\ttfamily
  astro-ph/9507009}}].

\bibitem{Raffelt:1987im}
G.~Raffelt and L.~Stodolsky, \emph{{Mixing of the Photon with Low Mass
  Particles}}, \href{https://doi.org/10.1103/PhysRevD.37.1237}{\emph{Phys.
  Rev.} {\bfseries D37} (1988) 1237}.

\bibitem{raffelt}
G.~G. {Raffelt}, \emph{{Particle Physics From Stars}},
  \href{https://doi.org/10.1146/annurev.nucl.49.1.163}{\emph{Annual Review of
  Nuclear and Particle Science} {\bfseries 49} (1999) 163}
  [\href{https://arxiv.org/abs/hep-ph/9903472}{{\ttfamily hep-ph/9903472}}].

\bibitem{raffeltbook}
G.~G. {Raffelt}, \emph{{Stars as laboratories for fundamental physics : the
  astrophysics of neutrinos, axions, and other weakly interacting particles}}.
  University of Chicago Press, 1996.

\bibitem{2012ApJ...757...14J}
R.~{Jansson} and G.~R. {Farrar}, \emph{{A New Model of the Galactic Magnetic
  Field}}, \href{https://doi.org/10.1088/0004-637X/757/1/14}{\emph{\apj}
  {\bfseries 757} (2012) 14} [\href{https://arxiv.org/abs/1204.3662}{{\ttfamily
  1204.3662}}].

\bibitem{2012ApJ...761L..11J}
R.~{Jansson} and G.~R. {Farrar}, \emph{{The Galactic Magnetic Field}},
  \href{https://doi.org/10.1088/2041-8205/761/1/L11}{\emph{\apj} {\bfseries
  761} (2012) L11} [\href{https://arxiv.org/abs/1210.7820}{{\ttfamily
  1210.7820}}].

\bibitem{Boehringer:2016kqe}
H.~Böhringer, G.~Chon and P.~P. Kronberg, \emph{{The Cosmic Large-Scale
  Structure in X-rays (CLASSIX) Cluster Survey I: Probing galaxy cluster
  magnetic fields with line of sight rotation measures}},
  \href{https://doi.org/10.1051/0004-6361/201628873}{\emph{Astron. Astrophys.}
  {\bfseries 596} (2016) A22}
  [\href{https://arxiv.org/abs/1610.02887}{{\ttfamily 1610.02887}}].

\bibitem{2013ApJS..208...20B}
C.~L. {Bennett}, D.~{Larson}, J.~L. {Weiland}, N.~{Jarosik}, G.~{Hinshaw},
  N.~{Odegard} et~al., \emph{{Nine-year Wilkinson Microwave Anisotropy Probe
  (WMAP) Observations: Final Maps and Results}},
  \href{https://doi.org/10.1088/0067-0049/208/2/20}{\emph{\apjs} {\bfseries
  208} (2013) 20} [\href{https://arxiv.org/abs/1212.5225}{{\ttfamily
  1212.5225}}].

\bibitem{2016A&A...594A...1P}
{Planck Collaboration}, R.~{Adam}, P.~A.~R. {Ade}, N.~{Aghanim}, Y.~{Akrami},
  M.~I.~R. {Alves} et~al., \emph{{Planck 2015 results. I. Overview of products
  and scientific results}},
  \href{https://doi.org/10.1051/0004-6361/201527101}{\emph{\aap} {\bfseries
  594} (2016) A1} [\href{https://arxiv.org/abs/1502.01582}{{\ttfamily
  1502.01582}}].

\bibitem{Eriksen:2005dr}
H.~K. Eriksen et~al., \emph{{CMB component separation by parameter
  estimation}}, \href{https://doi.org/10.1086/500499}{\emph{Astrophys. J.}
  {\bfseries 641} (2006) 665}
  [\href{https://arxiv.org/abs/astro-ph/0508268}{{\ttfamily
  astro-ph/0508268}}].

\bibitem{Delabrouille:2008qd}
J.~Delabrouille, J.~F. Cardoso, M.~L. Jeune, M.~Betoule, G.~Fay and
  F.~Guilloux, \emph{{A full sky, low foreground, high resolution CMB map from
  WMAP}}, \href{https://doi.org/10.1051/0004-6361:200810514}{\emph{Astron.
  Astrophys.} {\bfseries 493} (2009) 835}
  [\href{https://arxiv.org/abs/0807.0773}{{\ttfamily 0807.0773}}].

\bibitem{Rogers:2016rdb}
K.~K. Rogers, H.~V. Peiris, B.~Leistedt, J.~D. McEwen and A.~Pontzen,
  \emph{{Spin-SILC: CMB polarization component separation with spin wavelets}},
  \href{https://doi.org/10.1093/mnras/stw2128}{\emph{Mon. Not. Roy. Astron.
  Soc.} {\bfseries 463} (2016) 2310}
  [\href{https://arxiv.org/abs/1605.01417}{{\ttfamily 1605.01417}}].

\bibitem{Delabrouille:2002kz}
J.~Delabrouille, J.~F. Cardoso and G.~Patanchon, \emph{{Multi-detector
  multi-component spectral matching and applications for CMB data analysis}},
  \href{https://doi.org/10.1111/j.1365-2966.2003.07069.x}{\emph{Mon. Not. Roy.
  Astron. Soc.} {\bfseries 346} (2003) 1089}
  [\href{https://arxiv.org/abs/astro-ph/0211504}{{\ttfamily
  astro-ph/0211504}}].

\bibitem{Vansyngel:2014dfa}
F.~Vansyngel, B.~D. Wandelt, J.-F. Cardoso and K.~Benabed, \emph{{Semi-blind
  Bayesian inference of CMB map and power spectrum}},
  \href{https://doi.org/10.1051/0004-6361/201424890}{\emph{Astron. Astrophys.}
  {\bfseries 588} (2016) A113}
  [\href{https://arxiv.org/abs/1409.0858}{{\ttfamily 1409.0858}}].

\bibitem{Eriksen:2004jg}
H.~K. Eriksen, A.~J. Banday, K.~M. Gorski and P.~B. Lilje, \emph{{Foreground
  removal by an internal linear combination method: Limitations and
  implications}}, \href{https://doi.org/10.1086/422807}{\emph{Astrophys. J.}
  {\bfseries 612} (2004) 633}
  [\href{https://arxiv.org/abs/astro-ph/0403098}{{\ttfamily
  astro-ph/0403098}}].

\bibitem{Hurier:2010wu}
G.~Hurier, S.~R. Hildebrandt and J.~F. Macias-Perez, \emph{{MILCA: A Maximum
  Internal Linear Component Analysis for the extraction of spectral
  emissions}}, \href{https://doi.org/10.1051/0004-6361/201321891}{\emph{Astron.
  Astrophys.} {\bfseries 558} (2013) A118}
  [\href{https://arxiv.org/abs/1007.1149}{{\ttfamily 1007.1149}}].

\bibitem{2011MNRAS.410.2481R}
M.~{Remazeilles}, J.~{Delabrouille} and J.-F. {Cardoso}, \emph{{CMB and SZ
  effect separation with constrained Internal Linear Combinations}},
  \href{https://doi.org/10.1111/j.1365-2966.2010.17624.x}{\emph{\mnras}
  {\bfseries 410} (2011) 2481}
  [\href{https://arxiv.org/abs/1006.5599}{{\ttfamily 1006.5599}}].

\bibitem{Khatri:2014sra}
R.~Khatri, \emph{{Linearized iterative least-squares (LIL): a parameter-fitting
  algorithm for component separation in multifrequency cosmic microwave
  background experiments such as Planck}},
  \href{https://doi.org/10.1093/mnras/stv1167}{\emph{Mon. Not. Roy. Astron.
  Soc.} {\bfseries 451} (2015) 3321}
  [\href{https://arxiv.org/abs/1410.7396}{{\ttfamily 1410.7396}}].

\bibitem{Khatri:2018fjk}
R.~Khatri, \emph{{Data driven foreground clustering approach to component
  separation in multifrequency CMB experiments: A new Planck CMB map}},
  \href{https://arxiv.org/abs/1808.05224}{{\ttfamily 1808.05224}}.

\bibitem{2003ApJS..148....1B}
C.~L. {Bennett}, M.~{Halpern}, G.~{Hinshaw}, N.~{Jarosik}, A.~{Kogut},
  M.~{Limon} et~al., \emph{{First-Year Wilkinson Microwave Anisotropy Probe
  (WMAP) Observations: Preliminary Maps and Basic Results}},
  \href{https://doi.org/10.1086/377253}{\emph{\apjs} {\bfseries 148} (2003) 1}.

\bibitem{mask}
R.~{Khatri}. \url{http://theory.tifr.res.in/~khatri/muresults/}, 2015.

\bibitem{Adam:2015vua}
{\scshape Planck} collaboration, \emph{{Planck 2015 results. VIII. High
  Frequency Instrument data processing: Calibration and maps}},
  \href{https://doi.org/10.1051/0004-6361/201525820}{\emph{Astron. Astrophys.}
  {\bfseries 594} (2016) A8}
  [\href{https://arxiv.org/abs/1502.01587}{{\ttfamily 1502.01587}}].

\bibitem{pla}
{Planck Collaboration}, ``{Planck Legacy Archive}.''
  \url{http://pla.esac.esa.int/pla}, 2015.

\bibitem{2014A&A...571A...9P}
{Planck Collaboration}, P.~A.~R. {Ade}, N.~{Aghanim}, C.~{Armitage-Caplan},
  M.~{Arnaud}, M.~{Ashdown} et~al., \emph{{Planck 2013 results. IX. HFI
  spectral response}},
  \href{https://doi.org/10.1051/0004-6361/201321531}{\emph{\aap} {\bfseries
  571} (2014) A9} [\href{https://arxiv.org/abs/1303.5070}{{\ttfamily
  1303.5070}}].

\bibitem{2004MNRAS.349..603E}
G.~{Efstathiou}, \emph{{Myths and truths concerning estimation of power
  spectra: the case for a hybrid estimator}},
  \href{https://doi.org/10.1111/j.1365-2966.2004.07530.x}{\emph{\mnras}
  {\bfseries 349} (2004) 603}
  [\href{https://arxiv.org/abs/astro-ph/0307515}{{\ttfamily
  astro-ph/0307515}}].

\bibitem{polspice}
A.~{Challinor}, G.~{Chon}, S.~{Colombi}, E.~{Hivon}, S.~{Prunet} and
  I.~{Szapudi}, ``{PolSpice}.''
  \url{http://www2.iap.fr/users/hivon/software/PolSpice/}.

\bibitem{Hivon:2001jp}
E.~Hivon, K.~M. Gorski, C.~B. Netterfield, B.~P. Crill, S.~Prunet and
  F.~Hansen, \emph{{Master of the cosmic microwave background anisotropy power
  spectrum: a fast method for statistical analysis of large and complex cosmic
  microwave background data sets}},
  \href{https://doi.org/10.1086/338126}{\emph{Astrophys. J.} {\bfseries 567}
  (2002) 2} [\href{https://arxiv.org/abs/astro-ph/0105302}{{\ttfamily
  astro-ph/0105302}}].

\bibitem{2001ApJ...548L.115S}
I.~{Szapudi}, S.~{Prunet}, D.~{Pogosyan}, A.~S. {Szalay} and J.~R. {Bond},
  \emph{{Fast Cosmic Microwave Background Analyses via Correlation Functions}},
  \href{https://doi.org/10.1086/319105}{\emph{\apjl} {\bfseries 548} (2001)
  L115}.

\bibitem{2004MNRAS.350..914C}
G.~{Chon}, A.~{Challinor}, S.~{Prunet}, E.~{Hivon} and I.~{Szapudi},
  \emph{{Fast estimation of polarization power spectra using correlation
  functions}},
  \href{https://doi.org/10.1111/j.1365-2966.2004.07737.x}{\emph{\mnras}
  {\bfseries 350} (2004) 914}
  [\href{https://arxiv.org/abs/astro-ph/0303414}{{\ttfamily
  astro-ph/0303414}}].

\bibitem{2018arXiv180804493G}
N.~{Galitzki}, A.~{Ali}, K.~S. {Arnold}, P.~C. {Ashton}, J.~E. {Austermann},
  C.~{Baccigalupi} et~al., \emph{{The Simons Observatory: Instrument
  Overview}}, {\emph{ArXiv e-prints} (2018) }
  [\href{https://arxiv.org/abs/1808.04493}{{\ttfamily 1808.04493}}].

\bibitem{2016JLTP..184..805S}
A.~{Suzuki}, P.~{Ade}, Y.~{Akiba}, C.~{Aleman}, K.~{Arnold}, C.~{Baccigalupi}
  et~al., \emph{{The Polarbear-2 and the Simons Array Experiments}},
  \href{https://doi.org/10.1007/s10909-015-1425-4}{\emph{Journal of Low
  Temperature Physics} {\bfseries 184} (2016) 805}
  [\href{https://arxiv.org/abs/1512.07299}{{\ttfamily 1512.07299}}].

\bibitem{2016ApJS..227...21T}
R.~J. {Thornton}, P.~A.~R. {Ade}, S.~{Aiola}, F.~E. {Angil{\`e}}, M.~{Amiri},
  J.~A. {Beall} et~al., \emph{{The Atacama Cosmology Telescope: The
  Polarization-sensitive ACTPol Instrument}},
  \href{https://doi.org/10.3847/1538-4365/227/2/21}{\emph{\apjs} {\bfseries
  227} (2016) 21} [\href{https://arxiv.org/abs/1605.06569}{{\ttfamily
  1605.06569}}].

\bibitem{Abazajian:2016yjj}
{\scshape CMB-S4} collaboration, \emph{{CMB-S4 Science Book, First Edition}},
  \href{https://arxiv.org/abs/1610.02743}{{\ttfamily 1610.02743}}.

\bibitem{Matsumura:2016sri}
T.~Matsumura et~al., \emph{{LiteBIRD: Mission Overview and Focal Plane
  Layout}}, \href{https://doi.org/10.1007/s10909-016-1542-8}{\emph{J. Low.
  Temp. Phys.} {\bfseries 184} (2016) 824}.

\bibitem{2014AAS...22343901K}
A.~J. {Kogut}, D.~T. {Chuss}, J.~L. {Dotson}, E.~{Dwek}, D.~J. {Fixsen},
  M.~{Halpern} et~al., \emph{{The Primordial Inflation Explorer (PIXIE)}},  in
  \emph{American Astronomical Society Meeting Abstracts \#223}, vol.~223 of
  \emph{American Astronomical Society Meeting Abstracts}, p.~439.01, Jan.,
  2014.

\bibitem{2018arXiv180801369Y}
K.~{Young}, M.~{Alvarez}, N.~{Battaglia}, J.~{Bock}, J.~{Borrill}, D.~{Chuss}
  et~al., \emph{{Optical Design of PICO, a Concept for a Space Mission to Probe
  Inflation and Cosmic Origins}}, {\emph{ArXiv e-prints} (2018) }
  [\href{https://arxiv.org/abs/1808.01369}{{\ttfamily 1808.01369}}].

\end{thebibliography}\endgroup
\bibliographystyle{JHEP}
\end{document}